\documentclass[10pt,english]{article}
\usepackage[LGR,T1]{fontenc}
\usepackage[utf8]{inputenc} 

\usepackage{xcolor}
\definecolor{cblue}{HTML}{4e79a5}
\definecolor{corange}{HTML}{f18f3b}
\definecolor{cred}{HTML}{e0585b}
\definecolor{cgreen}{HTML}{77b7b2}
\definecolor{cgreen2}{HTML}{5aa155}
\definecolor{cgold}{HTML}{EDC958}
\definecolor{cpurple}{HTML}{af7aa0}
\definecolor{cbrown}{HTML}{9c7561}

\definecolor{black32}{HTML}{000000}
\colorlet{black32alpha}{black32!32}

\usepackage{titlesec}
\titleformat{\section}
  {\centering\bf\scshape}{\thesection}{0.5em}{}
\titleformat{\subsection}
  {\normalsize\bf}{\thesubsection.}{0.25em}{}[\vspace{-0.75em}]

\newcommand{\printAndAddSectionToC}[1]{%
    \if\relax\detokenize{#1}\relax\else\addcontentsline{toc}{section}{#1}#1\fi}
\newcommand{\printAndAddSectionToCWhichIsSubsection}[1]{%
    \if\relax\detokenize{#1}\relax\else\addcontentsline{toc}{subsection}{#1}#1\fi}
\newcommand{\printAndAddSectionToCWhichIsSubsectionTwo}[1]{%
    \bookmarksetup{startatroot}\if\relax\detokenize{#1}\relax\else\addcontentsline{toc}{subsection}{#1}#1\fi}
\titleformat{name=\section,numberless}
  {\centering\bf\scshape}{}{0pt}{\printAndAddSectionToC}

\titleformat{name=\subsection,numberless}
  {\normalsize\bf}{}{0pt}{\printAndAddSectionToCWhichIsSubsectionTwo}[\vspace{-0.75em}]

\usepackage{geometry}
\usepackage{setspace}
\usepackage[natbib, style=apa, uniquelist=false, uniquename=false]{biblatex}
\letbibmacro{orig-doi+url}{doi+url}
\renewbibmacro*{doi+url}{%
	\iffieldundef{doi}%
	{\iffieldundef{url}%
		{\printfield{howpublished}}%
		{\usebibmacro{orig-doi+url}}%
	}%
	{\usebibmacro{orig-doi+url}}%
}%

\AtEveryBibitem{\clearfield{month}}
\AtEveryCitekey{\clearfield{month}}
\AtEveryBibitem{\clearfield{labelmonth}}
\AtEveryCitekey{\clearfield{labelmonth}}
\AtEveryBibitem{\clearfield{day}}
\AtEveryCitekey{\clearfield{day}}
\AtEveryBibitem{\clearfield{labelday}}
\AtEveryCitekey{\clearfield{labelday}}

\DeclareBibliographyDriver{customa}{%
	\printfield{usera}%
	\finentry
}

\usepackage{nth}

\usepackage{comment}
\usepackage{color}
\usepackage{float}
\usepackage[position=top]{subfig}
\usepackage{amsmath}
\usepackage{amsthm}
\let\amsnewtheorem\newtheorem %
\usepackage{amsfonts}
\usepackage{mathrsfs}  %
\usepackage{amssymb}
\usepackage{stmaryrd}
\usepackage{graphicx}
\usepackage{appendix}

\usepackage[normalem]{ulem}   %

\usepackage{sepfootnotes}

\usepackage{bm}
\usepackage{enumitem}

\usepackage[ruled, algosection]{algorithm2e}
\usepackage{booktabs} %
\usepackage{tablefootnote}  %
\usepackage{makecell} %

\let\originalleft\left
\let\originalright\right
\renewcommand{\left}{\mathopen{}\mathclose\bgroup\originalleft}
\renewcommand{\right}{\aftergroup\egroup\originalright}

\usepackage{mathtools} 
\DeclarePairedDelimiter\abs{\lvert}{\rvert}%
\DeclarePairedDelimiter\norm{\lVert}{\rVert}%
\makeatletter
\let\oldabs\abs
\def\abs{\@ifstar{\oldabs}{\oldabs*}}
\let\oldnorm\norm
\def\norm{\@ifstar{\oldnorm}{\oldnorm*}}
\makeatother

\usepackage{todonotes}

\newcommand{\X}{\mathcal{X}}

\newcommand{\T}{\mathcal T}

\newcommand{\sfP}{\mathsf P}

\newcommand{\sfPx}{\mathsf P\hspace{-0.07em}_{\xd}}

\newcommand{\sfPxdash}{\mathsf P\hspace{-0.07em}_{\xd'}}

\newcommand{\sfQ}{\mathsf Q}

\newcommand{\Xcef}{\mathcal X_{\mathrm{CEF}}}

\usepackage{bbm}

\usepackage{multirow, colortbl}
\usepackage{algorithmic}
\usepackage{xpatch}

\newcommand{\xd}{\bm{x}}

\newcommand{\Multd}{D_{{\normalfont \textsc{Mult}}}^\delta}

\newcommand{\dX}{d_{\mathcal X}}

\newcommand{\dT}{\dPr}
\newcommand{\dPr}{D_{\Pr}}

\newcommand{\scD}{\mathscr{D}}
\newcommand{\D}{\mathcal{D}}

\newcommand{\DPSpecCol}{$\epsilon_{\color{cbrown}\mathcal D}$-DP\DPFlavCol}

\newcommand{\DPFlavCol}{$({\color{cgreen2}\mathcal X},\allowbreak {\color{cbrown}\mathscr D},\allowbreak {\color{cred}\dX},\allowbreak {\color{cgreen}\dPr})$}

\renewcommand{\epsilon}{\varepsilon}

\theoremstyle{definition}

\theoremstyle{remark}

\amsnewtheorem*{excont}{Example \continuation} %
\newcommand{\continuation}{??}

\SetKw{CONTINUE}{continue}
\SetKw{GOTO}{go to}
\SetKw{OUTPUT}{Output:}
\makeatletter
\xpatchcmd\algorithmic
{\newcommand{\IF}}
{%
	\newcommand\SCOPE{\begin{ALC@g}}%
		\newcommand\ENDSCOPE{\end{ALC@g}}%
	\newcommand{\IF}%
}
{}{\fail}
\makeatother

\usepackage{numprint}
\npthousandsep{,}

\usepackage[unicode=true,bookmarks=true,bookmarksnumbered=true,bookmarksopen=true,breaklinks=true,pdfborder={0 0 0},pdfborderstyle={},colorlinks=true,allcolors=blue,hypertexnames=false]{hyperref}
\hypersetup{pdftitle={refreshment-III}, pdfauthor={JBRGXLM}}
\usepackage{bookmark}

\title{Differential Privacy Meets Invariant Statistics: \\ Some Conundrums in Quantifying Trade-Offs\footnote{Forthcoming in \fullcite{hotzDataPrivacyProtection2026}}}

\author{James Bailie,\hspace{-0.1em}\footnote{bailie@chalmers.se, \\  \indent\ \ Department of Statistics, Harvard University, Cambridge, USA.}\ \ Ruobin Gong\footnote{ruobin.gong@rutgers.edu, \\ \indent\ \ Department of Statistics, Rutgers University, Piscataway, USA.}\ \ and Xiao-Li Meng\footnote{meng@stat.harvard.edu, \\  \indent\ \ Department of Statistics, Harvard University, Cambridge, USA.}}

\date{July 8, 2026} 

\begin{document}

\maketitle

\begin{abstract}
    This work was inspired by the question of whether data swapping, a popular form of statistical disclosure control used to protect many data products including three recent US Decennial Censuses, can satisfy differential privacy (DP). Given the existence of more than 200 formulations of DP (and counting), as a precondition to answering this question one must precisely specify what it actually means to be DP. Motivated by this observation, we first conduct a theoretical investigation into DP's fundamental essence, resulting in a five-building-block system explicating the \textit{who, where, what, how} and \textit{how much} aspects of DP. Instantiating this system in the context of the US Decennial Census, we then demonstrate the broad applicability and relevance of DP by comparing a swapping strategy like that used in 2010 with the TopDown Algorithm---the main DP method adopted in the 2020 Census. This chapter provides nontechnical summaries of these two pieces of work (developed elsewhere), as well as extended discussions on a number of issues they unearth that complicate the formulation and the navigation of the so-called privacy--utility trade-off: How can greater awareness of the five building blocks thwart privacy theatrics? How can invariants (statistics that are released as is, without any privacy protection) align with DP's philosophy of relative privacy? How do our results bridging traditional statistical disclosure control and DP allow a data custodian to reap the benefits of both these fields? And how can removing the implicit reliance on aleatoric uncertainty lead to new generalizations of DP? Our ultimate goal with these discussions is to deepen the theoretical basis, broaden the practical applicability, and reduce the misperception of DP---all without shaking its core foundations. 

\end{abstract}

\newpage
\section{What Motivated This Study?}
\label{secIntro}

Since its invention two decades ago by %
\citet{dwork2006calibrating}, %
differential privacy (DP) has received a tremendous amount of attention in research and practice. As a field, it aims to provide a tractable, mathematical framework to quantify and operationalize the evasive concept of privacy within the context of sharing (i.e., releasing) statistical data. Yet, driven by a myriad of constraints (e.g., challenges in establishing theoretical guarantees or in practical implementation), attempts to alter, enhance and relax the original, so-called `pure' DP definition of \citet{dwork2006calibrating} have led to a plethora of `impure' DP formulations. %
As such, the term DP today encompasses %
a broad class of technical standards conceptualizing what it means for a data release algorithm to be `private.' (See \cite{desfontaines2020sok}, for a dizzying, but still only partial, enumeration of this class.)%

From a practical perspective, the interest in DP---and the subsequent explosion of different formulations---is easy to understand. 
Privacy concerns
have increased substantially as our society adventures deeper into the digital age. Organizations in all sectors and at all levels are compelled to expend effort addressing 
the issue of data privacy, whether for noble reasons or for fear of liability; yet each entity is faced with its own unique set of concerns, necessitating its own custom solution \citep{schneiderDataAnonymizationHasnt2025}. The adoption of a form of DP by the United States Census Bureau (USCB) for its 2020 Decennial Census of Population and Housing is a shining example of organizational effort---one that has generated much theoretical contemplation and methodological advances, as well as controversies and emotions ranging from excitement to frustration (see the special issue of the \textit{Harvard Data Science Review}, ``Differential Privacy for the 2020 US Census: Can We Make Data Both Private and Useful?'' \cite{HDSRSPTwo}). %

Indeed, this chapter owes its existence to the bureau's adoption of DP. Because data privacy has been a central concern of the census %
for well over a century, seeing the USCB's recent development of the DP-inspired TopDown Algorithm (TDA) \citep{abowd2022topdown} for its 2020 Census, one naturally may wonder in what ways it improves upon their %
past methods for statistical disclosure control (SDC). %
In particular, the data swapping strategy used in 2010, just like the TDA, involves injecting artificial randomness into the published census tables. %
So could it satisfy a form of DP as well, which would then make it easier to compare both methods within a single, unified framework?

Those who adhere to the definition of pure DP may immediately declare that any form of swapping cannot satisfy DP because it leaves some aggregations of the data unaltered, and hence some statistics%
---termed its \textit{invariants}---will be released without any artificial noise injected.  By the same logic, the TDA also cannot satisfy DP because it is designed to maintain various total counts (e.g., state population counts) in order to comply with mandates set by the US Constitution and other external policy considerations. Yet such a strict adherence to a narrow perspective not only greatly limits the applicability of DP, but also fundamentally misperceives its essence. 
After all, DP is not concerned with protecting absolute privacy---no data release method can \citep{kiferNoFreeLunch2011}. Rather, it concerns the protection of information that can be revealed by an individual's confidential data but is otherwise unavailable. %
The constitutional and policy mandates reveal information that cannot be protected, and hence any adoption of DP---or any other data protection standard for that matter---must take that into account. (Besides which, the narrow perspective that DP does not allow any invariants is misplaced. DP has in fact accommodated some invariants from its very inception: the number of records in the confidential dataset can be released exactly under many DP formulations, including pure DP.) 

For the case of swapping, the matter becomes muddier as its invariants are not externally mandated but instead are inherent to its design. %
One can see immediately the potential complications with, and debates about, different designs and their resulting invariants. Confusions may easily result from comparing SDC methods %
that are based on different postulates of what is already known or considered not to need protection, akin to the trouble of comparing two distributions when they are conditioned on different variables. %

Such a problem is only one of many nuanced issues we have had to deal with as we seek precise and contextual answers to the question of whether swapping is differentially private. This question served as an impetus for a deep study of DP to reveal its statistical essence and to contemplate its practical complications.  Consequently, it should come as little surprise that our investigation took considerably more time, and pages, than initially expected. In fact, in addition to this chapter, it resulted in two other articles---``Privacy Differentials in Differential Privacy'' \citep[referred to herein as Part~I]{bailieRefreshmentStirredNot2024a} and ``Invariant-Preserving Deployments of Differential Privacy for the US Decennial Census'' \citep[referred to herein as Part~II]{bailieRefreshmentStirredNot2024b}. %
The present work will first provide an intuitive summary and explanation of these articles before elaborating on their broader and deeper implications, and discussing 
some of the %
subtle issues that tend to invite misunderstanding, misuse, and misplaced expectations of DP.  

More specifically, through five `questions of protection,' Section~\ref{secBuildingBlocks} provides a nontechnical introduction to the system of DP specifications developed in Part~I. To demonstrate the necessity of all five of these questions, Section~\ref{sectionSpendLessPLB} gives examples of how one might `game' DP by trading off the strength of one's answers to some of the questions with weaker answers to the other questions. Of course, we are not endorsing these examples; rather we use them to highlight possible `privacy theatrics' that can be exploited in a mathematical and seemingly objective framework such as DP. Section~\ref{secCensus} is an accessible overview of Part~II's main results, describing the evolution of the US Census's SDC protection. In particular, with Part~II's DP specification for a particular type of data swapping---which is similar to that used in 2010---Section~\ref{secCensus} makes comparisons between the censuses in 2010 and 2020 using DP-based descriptions of their SDC protection.

Our DP specification for data swapping serves as a bridge between traditional SDC and DP. It also highlights certain differences between the design philosophies of these two approaches, which in our view explain their distinct emphasis on different characteristics of a data protection method. The presence of these differences complicates the formulation and navigation of the ``privacy--utility'' (or protection--usability) trade-off. Section~\ref{secConundrums} discusses some of the new conundrums that our work brings to the fore in this regard.  For example, DP supplies mathematically rigorous guarantees and the composition property, whereas traditional SDC favors facial validity and logical consistency. In an effort to reap the benefits of both these approaches, how should the data custodian harmonize their distinct foci? 

Another tension arises between DP's principle of `transparency for free' and the notion of `privacy through obscurity' found in traditional SDC. In our view, this tension hinges on the two fields' divergent views on the role that epistemic uncertainty plays in privacy protection. This observation opens up new research directions---for example, extending existing DP definitions to account for epistemic uncertainty via imprecise probabilities so as to retain rigorous guarantees of protection while preserving a data user's ability to conduct valid statistical analysis.

Integrating Parts~I and~II with the present chapter, this trio of works reflects our triple %
ambition: firstly, to leverage the merits of DP, including its mathematical assurances of protection and algorithmic transparency without sidelining the advantages of classical SDC; secondly, to unveil the nuances and potential pitfalls in employing DP as a theoretical yardstick for SDC procedures; %
and thirdly, to build connections between social and mathematical conceptualizations of data privacy by outlining real-world considerations behind the five-building-blocks system developed in Part~I.

This chapter in particular focuses on conundrums in quantifying the trade-offs that are inherent to deploying DP. Of primary interest is the trade-off between invariant statistics and the standard way of measuring protection in DP, the ``privacy loss budget.'' As we will see, this trade-off will be central to our comparisons between data swapping and the 2020 Census (Section~\ref{secCensus}). More generally, Part~I's system of DP specifications reveals that one can trade off any of the five building blocks of DP with one another (Section~\ref{sectionSpendLessPLB}). While this chapter exposes these trade-offs and provides new views, or conundrums, on established trade-offs, additional work is needed in order to quantify the trade-offs' different components---a task that is made difficult by the fact that the effects of these components on privacy protection are not directly comparable, but which may nevertheless be amenable to an economic analysis. %
Therefore, our conundrums are double: the novel perspectives unearthed by Parts~I and~II, as well as the ensuing challenges in quantifying the resulting trade-offs.

We should note that these conundrums are fundamental to deploying DP and are not unique to our primary application area, the US Decennial Census. For one, invariant statistics appear in many other data products (see Part~II and references therein). Moreover, deploying DP does not only require choosing which statistics will be invariant, but also requires answering the question of which steps of the data pipeline will be taken as invariant. For an exposition of the complexities and trade-offs inherent to this question in the context of survey data, see the chapter by \citet{drechslerComplexitiesDifferentialPrivacy2024} in this volume.

\section{Highlights of Part I:  Five Building Blocks of DP}\label{secBuildingBlocks}

Every formulation of DP is a 
way to measure 
the `privacy' of a \textit{data release mechanism}, a function that transforms a confidential dataset into publishable statistics (also referred to as an SDC method, a data sharing algorithm, or similar).
As we will review shortly, the core idea behind all of these formulation is to quantify privacy---or rather loss of privacy---as the amount of change in a mechanism's likelihood function, relative to a corresponding change in the individual data points.\footnote{The likelihood function of a mechanism is the map from the mechanism's inputs to the mechanism's output---or more correctly, the probability distribution of the mechanism's output.} %
Indeed, this narrow, mathematical conceptualization of data privacy in terms of a rate of change highlights DP's central tenet that privacy is controlled by bounding the `derivative' of a mechanism, as is alluded to by its nomenclature `differential.'

A technically oriented reader may immediately ask a host of questions. On what space is the likelihood function defined?  How is relative change metricized? What constitutes an ``individual data point''? What does ``control'' mean? And so on. 
Indeed, these questions are key to formalizing the above intuitive idea of DP into a concrete, mathematical definition %
and their many possible answers are 
reflected in the numerous formulations of DP found in the literature.
However, while it may appear there are many questions to formalize, it turns out five are sufficient (and necessary) to fully characterize most existing definitions of DP---these are the five \textit{questions of protection}, each of which correspond to what we call a \textit{building block} of DP: the domain $\X$ (who?); the multiverse $\scD$ (where?); the input premetric $\dX$ (what?); the output premetric $\dPr$ (how?); and the protection loss budget $\epsilon_{\D}$ (how much?).
A different choice for any one of these building blocks---i.e., a different answer to any one of the questions of protection---results in a different formulation of DP. %
As defining DP requires answering all five questions, instantiations for each of the building blocks collectively form a \textit{DP specification}---a complete, self-contained formalization of the intuitive idea of DP as a bound on a mechanism's rate of change.

\subsection{Questions of Protection}

To explain how these five building blocks specify DP,
we start with the most basic question concerning a mechanism, ``\textit{\textbf{who} is eligible for protection?}'', which is addressed by specifying all the actual, potential or counterfactual datasets that could be inputted into the mechanism. %
The collection of all such datasets is called the data space, or the \textit{domain}, and is denoted by $\mathcal{X}$. Because setting down all these datasets requires situating the actual dataset $\xd$ in the context of its lifecycle (how else would one know what other, counterfactual datasets are possible?)---not just specifying the mathematical schema of $\xd$---the domain $\X$ provides the meaning of who $\xd$'s data subjects are and how $\xd$ represents them.

As briefly mentioned at the start of this section, an SDC method computes some output statistics based on the actual, confidential dataset $\xd$. (It is worthwhile to emphasize that in the context of DP, while SDC methods are typically random functions of $\xd$, the randomness is not `in' $\xd$, but rather it is artificially injected into the statistics output by the SDC methods, in order to reduce the information these statistics reveal about $\xd$.) 
From an attacker's perspective, the confidential dataset $\xd$---which always belongs to $\mathcal X$ by design---%
is the unknown `parameter' to be inferred from the output statistics, hence the choice of $\mathcal X$ is conceptually analogous to the choice of a parameter space in standard statistical inference.  %

Explicating $\mathcal{X}$, however, is only the first step.  The next question is ``\textit{\textbf{where} does the protection extend to?}'' which can be answered by specifying a \textit{multiverse} $\mathscr D$, which is a collection of the possible \textit{universes} $\mathcal D$, be they actual or hypothetical.  The need for---and the distinction between---the data multiverse and the actual data universe is well illustrated by the application of DP to the US Decennial Census. Suppose the enumerated US population size $N_{US}$ is 330 million. Then, in order to comply with the constitutional mandate that the state population totals must be released exactly as enumerated, any hypothetical census dataset that does not yield $N_{US}=\numprint{330000000}$ will not be within the scope of protection, since any attacker can easily rule out such datasets. In this instance, if $N_{US}$ is the only aggregation that must be disclosed as is, then the corresponding data universe is simply all datasets in $\mathcal{X}$ such that the corresponding $N_{US}=\numprint{330000000}$.  

However, few would call for designing a mechanism that works only when $N_{US}$ is exactly 330 million. 
The enumerated US total population count varies from census to census, and indeed it varies within each census due to corrective adjustments (e.g., for reducing the impact of under-counting).
While it is essential for any data release mechanism to respect constitutional mandates---i.e., to disclose the total state enumerations---the actual value of the disclosed $N_{US}$ is rather accidental.  A sensible design should work regardless of the value of $N_{US}$, at least for values within a reasonable range %
(e.g., a mechanism need not be required to work for implausible values such as $N_{US}=330$ billion). 
To do this, we use a data multiverse that collects all the possible data universes, where each data universe is a set of datasets with the same value for $N_{US}$.
This means that all datasets in $\X$ are protected---but only within the scope of their respective universes.

Next is the question, ``\textit{\textbf{what} is the granularity of protection?}'' %
which {can be} a source of confusion, as well as an opportunity for manipulation in capable but malicious hands. Recalling that a DP specification is a bound on a mechanism's rate of change, the granularity of protection informally corresponds to the entities whose data is counterfactually altered when measuring this rate of change. These entities are termed the \textit{protection units} (or, elsewhere in the literature, the privacy units).
Individual persons or business entities are common choices for the protection units, but they are not the only ones. 
For example, for a dataset of electronic communications, the protection unit may be defined as a single message sent by a person, rather than the sender herself. As an individual may send many messages a day, such a fine-grained protection unit allows a social media platform to declare a nominal level of DP protection that appears impressively high, %
even though the actual risks to the sender remain exponentially large.

Mathematically, the granularity of protection is formally conceptualized via the input premetric $\dX(\xd, \xd')$, which is a measure of the difference (or `distance') between two datasets $\xd$ and $\xd'$ in the domain $\mathcal X$. 
Furthermore, the protection units correspond conceptually to unit differences in $\dX$---i.e., the differences between \textit{neighboring} (or adjacent) datasets, which are pairs of counterfactual datasets $\xd,\xd'\in\X$ with $\dX(\xd, \xd') = 1$. 
(More exactly, a protection unit %
is what differs during the generating processes of two neighboring datasets.)
While neighboring datasets could differ by the deletion of one record, or the alteration of a single attribute, 
to properly define the protection units requires an appreciation of what a unit difference in $\dX$ actually represents in the real world. This in turn necessitates placing $\xd$ within the context of its data pipeline, highlighting once again the importance of the domain $\X$'s contextual definition over and above a purely mathematical description of it. %

As we %
have repeatedly described, DP quantifies loss of `privacy' in terms of the rate of change in the variations of a data release mechanism's output. For each specification of DP, this rate of change is calculated with respect to the specification's input premetric $\dX$, but \textit{\textbf{how} is the change in output variations measured}? A premetric is also used to address this question, specifically the specification's output premetric $\dPr$. While $\dX(\xd, \xd')$ measures differences between datasets $\xd$ and $\xd'$, the output premetric $\dPr(\sfPx, \sfPxdash)$ is a measure of the difference between probability distributions $\sfPx$ and $\sfPxdash$.
Here, $\sfPx$ denotes the likelihood function---i.e., the probability distribution of the {released statistics}, as a function of the confidential dataset $\xd$,  where the randomness in $\sfPx$ is introduced solely by the data release mechanism. This is a sensible approach to SDC: as all statistical information is created by variations, by limiting the difference between the output distributions $\sfPx$ and $\sfPxdash$, we limit an attacker's ability to distinguish between $\xd$ and $\xd'$. Furthermore, by controlling the rate of change, rather than the absolute change, we allow large differences between $\sfPx$ and $\sfPxdash$ when $\dX(\xd, \xd')$ is also large (intuitively meaning that the difference between $\xd$ and $\xd'$ is not considered confidential information); while still restricting $\dPr(\sfPx, \sfPxdash)$ when $\dX(\xd, \xd')$ is small.

This brings us to our fifth and final question, ``\textit{\textbf{how much} protection is afforded?}'' whose answer is given by the bound the DP specification imposes on a mechanism's rate of change. %
This bound is given by the building block $\epsilon(\cdot)$, a function mapping each universe $\D \in \scD$ to a non-negative (potentially infinite) value $\epsilon(\D)$. For notational convenience, we denote both the value $\epsilon(\D)$ and the function $\epsilon(\cdot)$ by $\epsilon_{\D}$ except when we need to explicitly differentiate between them.

For each universe $\D$, the quantity $\epsilon_{\D}$ is the specification's bound on the rate of change within the universe $\D$;
we say that a mechanism satisfies a DP specification if its rate of change in each universe $\D$ is bounded by the value $\epsilon_{\D}$. Smaller values of $\epsilon_{\D}$ are more restrictive and hence supply higher levels of SDC protection to $\D$ than larger values of $\epsilon_{\D}$. We allow the bound $\epsilon_\D$ to vary between universes so that different universes can be afforded different degrees of protection. This is desirable for example when, due to data utility concerns, the data custodian wants some entities to receive less protection than others; or when they would like to measure protection ``per-attribute'' \citep{seemanPrivatelyAnsweringQueries2023, ashmead2019effective}. 

We propose the term \textit{protection loss budget} for the function $\epsilon_{\mathcal D}$ instead of the more commonly used phrase ``privacy loss budget'' (although we maintain the same abbreviation, PLB). This reflects our desire to avoid running the risk of misrepresenting DP as a panacea for the diversity of data privacy issues found in modern society \citep{nissenbaumPrivacyContextTechnology2010}. In fact, DP only addresses a very narrow, but legitimate, concern \citep{seemanPrivacyUtilityDifferential2023}---%
an observation which also explains our use of quotation marks when discussing DP as a conceptualization of `privacy': While we strive to %
describe DP in a way that stays faithful to its literature, %
we %
also want to remain cognizant of the fact that, even in the specific context of statistical data sharing, there are dimensions to privacy that are outside the gamut of DP \citep{bailieFiveSafesPrivacyContext2023}.

\subsection{What Is a Specification of DP, Formally?}

Now, having described each of the five building blocks of a DP specification in turn---the domain $\X$, the multiverse $\scD$, the input premetric $\dX$, the output premetric $\dPr$ and the PLB $\epsilon_{\D}$---we can now formally tie them together with the following definition: A \textit{DP specification} %
is the condition on a data release mechanism that
\begin{equation}\label{eqSec2Derivertive}
\frac{\dPr(\sfPx, \sfPxdash)}{d_{\mathcal X}(\xd, \xd')} \le \epsilon_{\D}, \quad \text{or more generally,} \quad \dPr(\sfPx, \sfPxdash)\le \epsilon_{\D} d_{\mathcal X}(\xd, \xd'),
\end{equation}
for all $\xd, \xd' \in \mathcal D$ and all $\mathcal D\in \scD$. We provide two expressions here because the first one captures the intuition of DP as a bound on the mechanism's rate of change (i.e., the change in $\dPr$ per corresponding change in $\dX$); %
while the second shows that mathematically a DP specification is simply a Lipschitz continuity condition on $\sfPx$ as a function of the input data $\xd$. (A Lipschitz condition on a function is a generalization of the condition that the function's derivative is bounded.)

\begin{figure}
	\centering
    \includegraphics[width = \textwidth]{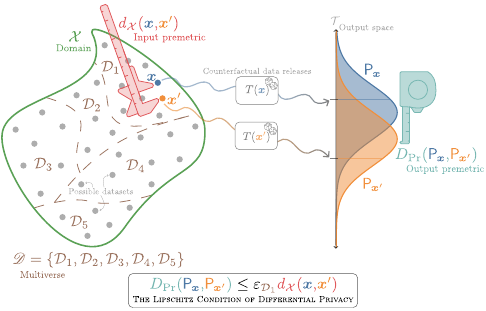}
	\caption{\textit{Schematic of a differential privacy specification} \DPSpecCol. The {\color{cgreen2}\textit{domain} $\X$} is the set of all possible datasets (be they actual, potential or counterfactual). We denote two arbitrary datasets by {\color{cblue} $\xd$} and {\color{corange} $\xd'$}; other possible datasets are depicted by {\color{black32alpha} gray circles}. The {\color{cbrown} \textit{multiverse} $\scD = \{ \D_1, \D_2, \D_3, \D_4, \D_5\}$} is a collection of sets of datasets---these sets are called \textit{universes}. (In this schematic, $\scD$ partitions the domain $\X$, as would happen when $\scD$ encodes invariants. In general, this need not be the case. In fact, often the universes may be overlapping.) 
	A data release mechanism $T$ transforms a dataset $\xd$ %
	to a random output $T(\xd)$, %
	which is a draw from the probability distribution $\sfPx$. %
	Intuitively, differential privacy requires that similar datasets {\color{cblue} $\xd$} and {\color{corange} $\xd'$} have similar \textit{output distributions} {\color{cblue}$\sfPx$} and {\color{corange}$\sfPxdash$}. This is formalized by the Lipschitz condition $\dPr(\sfPx, \sfPxdash) \le \varepsilon_{\D_1} \dX(\xd, \xd')$, 
	that states that the `distance' $\dPr(\sfPx, \sfPxdash)$ between the output distributions is at most a constant multiple $\varepsilon_{\D_1}$ of the `distance' $\dX(\xd, \xd')$ between the corresponding input datasets.
	Here, similarity (or `distance') between datasets %
	is measured by the DP specification's {\color{cred}\textit{input premetric}} ${\color{cred}\dX}$, %
	visualized above as a caliper, %
	and similarity %
	between probability distributions of the output under different inputs %
	is measured by the DP specification's {\color{cgreen}\textit{output premetric}} ${\color{cgreen}\dPr}$ %
	(the tape measure). %
	For simplicity, we depict the output space $\T$ as one dimensional, although in practice it is frequently a high-dimensional space, or even a union of many different probability spaces (as in local DP). (The PLB above, $\varepsilon_{\D_1}$, has the subscript $\D_1$ because the Lipschitz condition is applied to the datasets $\xd$ and $\xd'$ that are members of the universe $\D_1$, and because the PLB is allowed to vary between universes, potentially taking up to five different values, $\varepsilon_{\D_1}, \varepsilon_{\D_2}, \ldots, \varepsilon_{\D_5}$---one for each of the five universes.)}
	\label{figDPDiagram}
\end{figure}

A DP specification is characterized by its choices for the five building blocks, with different choices generating different DP specifications. While similar ideas already exist (some of which directly inspired this work), the usefulness of our system of DP specifications comes from the fact that: 1) it is a sufficiently general theory to encompass most published formulations of DP, thereby providing a unifying framework for a vast swathe of literature; and yet 2) each DP specification is a complete, mathematically precise definition of DP. %
In contrast, 
many common notions of DP are concerned with only one building block, leaving the others unspecified, and hence are not well-defined formulations of DP. For example, pure $\epsilon$-DP, approximate $(\epsilon, \delta)$-DP, and $\rho$-zero concentrated DP ($\rho$-zCDP) specify choices for the output premetric $\dPr$ only. This is evidenced by the fact that there are both bounded and unbounded versions of these three notions---and yet the bounded-vs-unbounded distinction itself only partly specifies the choice of $\dX$ (as either a Hamming or symmetric difference distance) since it still leaves unspecified the protection unit of $\dX$ (i.e., at what granularity the Hamming or symmetric difference distance is defined). 
Even after a data custodian's choices for all of the above have been clarified, %
an observer still does not have a complete understanding of the custodian's formulation of DP (since they do not know what form of the data is even being protected)---a confusing situation which warrants systematization.
Our aim with Part~I's system of DP specifications is to resolve this confusion.

In summary, a DP specification is a technical standard against which a data release mechanism can be assessed (either a mechanism satisfies the condition given in Equation~\ref{eqSec2Derivertive}---in which case it meets the DP specification---or it does not satisfy Equation~\ref{eqSec2Derivertive} and hence does not meet the DP specification); and a DP specification is defined by its five components (or building blocks), which are:

\begin{enumerate}
    \item The \textit{domain} $\X$: the set of all actual, potential or counterfactual datasets to be protected under the DP specification. Intuitively speaking, $\X$ describes the `domain of protection' and answers the question, ``\textit{\textbf{who}} is eligible for protection?''
    \item The \textit{multiverse} $\scD$: the collection of the DP specification's universes, which are subsets of $\X$ and collectively instantiate the `scope of protection' provided by the specification. $\scD$ answers the question, ``\textit{\textbf{where}} does the protection extend to?''
    \item The \textit{input premetric} $\dX$: the DP specification's measure of difference (or `distance', loosely speaking) between any two datasets $\xd$ and $\xd'$ in the domain $\X$. $\dX$ conceptualizes the `protection unit' and answers the question, ``\textit{\textbf{what}} is the granularity of protection?''
    \item The \textit{output premetric} $\dPr$: the DP specification's measure of difference between any two of the released data's possible probability distributions (e.g., $\sfPx$ and $\sfPxdash$). $\dPr$ captures the specification's `standard of protection,' answering the question, ``\textit{\textbf{how}} are changes in output variations measured?''
    \item The \textit{protection loss budget} $\epsilon_{\D}$: 
    the function $\epsilon_{\D}$ which supplies a value $\epsilon_{\D}$ for each universe $\D \in \scD$, quantifying the DP specification's `intensity of protection' and answering the question, ``\textit{\textbf{how much}} protection is afforded?''
\end{enumerate}

Much of the literature on DP treat the PLB as the main 
measure of the strength of a DP formulation. For example, \citet{abowd2016economic} uses it as the sole quantity for balancing the trade-off between privacy and utility. %
However, the five-building-blocks system reveals that the PLB can only be meaningfully defined after the \textit{DP flavor}---that is, the collection of the first four building blocks---has been fully spelled out. %
That is to say, only answers to `Who?,' `Where?,' `What?' and `How?' can make the answer to `How much?' concrete. 
In fact, a DP flavor can be thought of as a yardstick measuring the protection given by a data release mechanism, while the PLB is a specific measurement on that yardstick; and choices for both the yardstick and the measurement can be used when finding the optimal privacy--utility trade-off.
To make another analogy, PLBs are to their DP flavors as banknotes are to their currencies. At the end of the day, it is not the number on the banknote (or the value of the PLB) that matters, but its purchasing power---which is determined by the strength of the note's currency (or the PLB's DP flavor).
For example, if Japan announces that tomorrow their currency will be denominated by ``centiyen'' (which is equal to one hundred yen), the number in a Japanese person's bank account would change, but the number of Macbooks they could afford would not.
In the same way, a `redenomination' of the data custodian's DP flavor will change their PLB without affecting the actual level of SDC protection, as we explore in the next section.
Moreover, this analogy reminds us that just like currencies are proxies to purchasing power, determined by their complex economic and political environment, %
DP flavors are proxies to the contextual, fluid concept of privacy.%

\section{How to Reduce `Privacy Loss' Without Adding More Noise: A Perverse Guide}\label{sectionSpendLessPLB}

One lesson from the previous section %
is that the PLB is deficient as an unqualified measurement of %
SDC protection. Any description %
that omits the other building blocks opens the door to a variety of 
ways to `cheat'---i.e., ways to reduce the PLB without adding more noise. 
While these ways are not merely mathematical possibilities or pathological cases, but rather consequences of 
taking the value of $\epsilon_{\mathcal D}$ nominally and out of context, their mathematical feasibility can be explained rather simply: In order to alter one component of a DP specification---in this case the PLB---%
one can maneuver the other components ($\X$, $\scD$, $\dX$ or $\dT$), or the parameters of the data release mechanism, to that end. %
These maneuvers may be valid, particularly when the data custodian is transparent about the resulting DP specification, but they can also be used to misleadingly promote a high nominal level of protection (small PLB) especially if shortcomings in the other building blocks are deliberately hidden. 
Needless to say, our intention is not to encourage unscrupulous behavior, but rather to expose the inherent weaknesses that are open for exploitation in what might appear to be an objective and mathematically absolute framework for `privacy' protection. These warnings may be particularly relevant to commercial implementations of DP, where conflicts of interest are commonplace \citep[see e.g.][]{waldman2021industry}. For example, when the data custodian and data user are the same entity (such as a tech company collecting data about their customers), they may be tempted to cut some differentially private corners---or engage in some \textit{privacy theatrics} \citep{smartUnderstandingRisksPrivacy2022}---%
so as 
to simultaneously improve data utility while maintaining prima facie privacy protection to assuage their data contributors. %

\subsection{Cheating the System of DP Specifications}

In cataloging some of these ways to `cheat,' we start with the first ingredient of a DP flavor: the protection domain, $\mathcal X$.  The less to protect, the less protection budget needs to be spent. Therefore, choosing or interpreting  $\mathcal X$ more restrictively may lead to a smaller PLB, even without actually adding more noise to the data. This potentially perverse incentive should receive more scrutiny, because often the choice of $\mathcal X$ is a unconscious one on the data custodian's part---they are simply given the data they are given---but is indispensable to the interpretation of the DP specification. The domain $\mathcal X$ frames the variables that are contained in the confidential dataset $\xd$ and thus determines the socio-cultural sensitivity of what is being protected---an integral part of any understanding of privacy \citep{nissenbaumPrivacyContextTechnology2010}. More generally, every choice of $\mathcal X$  encodes a data conceptualization \citep{leonelliDataGovernanceKey2019}---a representation \textit{by} the data \textit{of} the individuals who contributed their information. Viewing the data release mechanism as a constituent phase in a data life cycle,  $\mathcal X$ specifies the starting point of that cycle. The impact of this choice permeates through other phases in the cycle, notably before data privatization including conceptualization, coding, cleaning, imputation, (sub-)sampling, and so on \citep{mengEnhancingPublicationsData2021}. Restrictions imposed by each of these steps may impact $\mathcal X$ before the privatization step, thus affecting the PLB \citep{huProvablePrivacyNonprivate2024a}. Take the concrete example of training a large machine learning model, which often involves an initial ``pretraining'' step, followed by ``fine-tuning.'' In this setting, some DP implementations take $\xd$ to be the data used for fine-tuning; by considering the pretrained model as given, any data used in pretraining is not included in the domain and hence not subject to DP protection \citep{tramerPositionConsiderationsDifferentially2024}. 
For another example, consider implementing DP in the context of a statistical survey, where the data release mechanism could plausibly start before or after the sampling step---a choice that would change the interpretation of what is being protected, as well as what assumptions about the attacker are required to ensure the nominal level of DP protection \citep{bailieWhoseDataIt2024}.

A second way to spend less PLB without adding more noise is to change the multiverse $\scD$, the second ingredient of a DP flavor. For example, piling on more invariants---i.e., summaries of data that will be published exactly by the data release mechanism---can reduce the PLB because it creates a more shattered data multiverse $\scD$. (A multiverse $\scD$ is a shattering, or refinement, of another multiverse $\scD'$ if every $\D \in \scD$ is a subset of some $\D' \in \scD'$.) For those who appreciate DP as a framework for protecting \textit{relative} privacy, this possibility is rather obvious.  The more one discloses via the invariants, the less information there is left in the data that requires protection, and hence a smaller PLB is incurred. Take swapping as a concrete case: as shown in Part~II and briefly recapped in Section~\ref{sec:dp2010}, the PLB $\epsilon$ of the Permutation Swapping Algorithm is determined by the swap rate $p$ and the largest stratum size $b$. To decrease the nominal value of $\epsilon$, one can either increase $p$ (up to a point) or decrease $b$. When the dataset has a fixed size, the simplest way to decrease $b$ is to define the stratifying variables at a finer resolution, resulting in smaller strata within which swapping is confined.  As illustrated in Part~II, the various choices of the stratifying variables at different levels of geography, with or without crossing with the household size variable, result in $b$ ranging from as small as 11.7 thousand to as large as 13.7 million, and a nominal $\epsilon$ from $12.31$ to  $19.38$ (respectively, when using a swap rate of $p = 5\%$). This shows how adding invariants can reduce the PLB.

A third way to achieve a nominal reduction of the PLB is to redefine the protection units---as captured by the third ingredient of a DP flavor, $d_{\cal X}$---to have a finer granularity. With all else being equal, a DP specification with coarser protection units packs more weight in its PLB; it offers a stronger protection guarantee at the same nominal budget than a specification with finer protection units. In the opposite direction, when a more expansive protection unit is supplanted by a narrower one, the input premetric $\dX$ becomes inflated: a unit change in the former sense may amount to multiple units of change in the latter sense, ``watering down'' the PLB by the same amount. 

This maneuver has been recognized, and to some extent utilized, in the literature on the design of data release mechanism for complex data structures. For example, the choice of neighbors is particularly important for network data: Are neighbors defined by removing a node or an edge from the network, that is, are protection units edges or nodes \citep{raskhodnikovaDifferentiallyPrivateAnalysis2016}? For business databases, does a company constitute a unit, or should units be employees, or both \citep{haneyUtilityCostFormal2017, schmutteDifferentiallyPrivatePublication2016, he2014blowfish}? Or should they be the company's transactions? Similarly for large personal databases in commercial settings, should an individual constitute a unit, or should each of their interactions with the platform (such as a post or a `like') be protection units, or should units be the set of a user's interactions within a given time period, such as a single day \citep{kenthapadiPriPeARLFrameworkPrivacyPreserving2018, DVN/TDOAPG/DGSAMS_2020, desfontainesListRealworldUses2023}? Finally, when publishing social statistics, do households deserve SDC protection above and beyond the protection afforded to their individual members \citep{machanavajjhalaCandidateDifferentialPrivacy2022}? Evidently, a data custodian often has a choice regarding which types of entities should be the protection units---a choice that can lead to meaningful protection, or to a superficial nominal guarantee that masks substantive vulnerabilities.

A fourth way to gain nominal PLB out of thin air is to artificially introduce an output premetric $\dT$ that systematically assesses two distributions to be closer than would otherwise be the case. Technically speaking, the relaxation from pure $\epsilon$-DP to $(\epsilon,\delta)$-approximate DP (ADP) \citep{dworkOurDataOurselves2006} can be understood as a maneuver of this type.\footnote{This is not to say that $(\epsilon,\delta)$-ADP is never a valid choice---in certain situations the gains to data utility may legitimately outweigh the loss to SDC of adopting $(\epsilon,\delta)$-DP. However, often there are other choices that allow for the same gains in utility while requiring that a data release mechanism `fail gracefully' rather than allowing a non-zero probability of `catastrophic failure' \citep[see e.g.][Chapter~7]{nearProgrammingDifferentialPrivacy}.} This can be seen by writing out the choice of output premetric corresponding to $(\epsilon,\delta)$-ADP (see Part~I):  
\begin{equation}\label{eq:delta}
\Multd (\sfP,\sfQ) = \sup_{S \in \mathscr F} \left\{ \left. \ln \frac{\left[\sfP(S)-\delta\right]^+}{\sfQ(S)}, \ln \frac{\left[\sfQ(S)-\delta\right]^+}{\sfP(S)} \right., 0 \right\},
\end{equation}
where $[x]^+ = \max\{x, 0\}$; $\sfP$ and $\sfQ$ are two probability measures on the same output space $\T$; and $\mathscr F$ is the $\sigma$-algebra on $\T$ (i.e., the collection of events on this output space that are of interest and that permit logically coherent probabilistic assignment). %
Clearly, for any $\delta >0$ and $\sfP \ne \sfQ$, we have that $\Multd (\sfP,\sfQ) <  \sup_{S \in \mathscr F} \left|\ln \sfP(S)-\ln \sfQ(S)]\right|$. Since the right hand side of this equation is the output premetric corresponding to pure $\epsilon$-DP, we have reduced the PLB without changing the actual data release mechanism.  %

\subsection{Other \texorpdfstring{$\epsilon$}{Epsilon} Evasion Tactics}

Because the PLB is used to bound the worst-case rate of change, as seen in Equation~\ref{eqSec2Derivertive}, any strategy that reduces the extremeness of the worst case will permit a smaller PLB without injecting any additional noise. %
For example, we can reduce the $\sigma$-algebra ${\mathscr F}$ by pretending that we are interested in fewer possible output events. %
This can be seen clearly in Equation~\ref{eq:delta}, where the right-hand side is the largest possible value over any possible event $S$ in ${\mathscr F}$. %
If we reduce $\mathscr F$ to a sub-$\sigma$-algebra, then the largest possible value may decrease, affording us with a smaller PLB without actually changing the behavior of the mechanism. 

The concept of subspace differential privacy \citep{gao2022subspace} reflects a maneuver of this type, as it requires control over the output only within sets in the Borel $\sigma$-algebra generated by a linear subspace of the ambient output space. Coarsening the $\sigma$-algebra associated with the output space signifies a weaker standard against which the data release mechanism is assessed, yet it does not compel the mechanism to be non-measurable with respect to another richer $\sigma$-algebra. Therefore, for subspace differential privacy, the mechanism could still take values outside the linear subspace, but no guarantees of SDC protection would be ensured by DP in such cases. (This is different from the requirement of invariants, which operates on the input space rather than the output space of the data release mechanism.)

To further emphasize the importance of understanding the vulnerabilities and subtleties inherent to DP, consider the case where the data custodian uses a constant data release mechanism---that is, a mechanism of the form $T(\xd)=C$, where $C$ does not change with $\xd$. Clearly such a data release mechanism has zero PLB regardless of the DP flavor under consideration, because it is completely insensitive to any manipulation of the input data $\xd$. But what if the data curator chooses $C$ to be the same value as the very data or query they are ostensibly trying to protect? Surely that means the PLB should be infinite, since the actual query or data is disclosed exactly. Whereas this may appear to be a pathological case, it carries a critical message: the design of a mechanism cannot be permitted to depend on the confidential dataset itself. This is the very reason that, when designing the 2020 DP mechanisms, the USCB used the 2010 Census data, rather the 2020 data \citep{uscensusbureauDevelopingDemonstrationData2023}. %

Yet the 2010 and 2020 Census data are highly correlated. Indeed, if a respondent's data did not change in 2020 as compared to 2010 (except obviously adding ten years to their age), and the attacker knew this fact, can we really say their data was protected under the DP specification of the 2020 Census? Not necessarily---in the worst case, the DP specification of any data release mechanism is %
conditional on the attacker knowing all of the data used in the development of the mechanism. This might prompt a data custodian to take a meta-perspective, where the data-dependent design of the mechanism is itself considered as part of a `meta-mechanism.'
For the constant mechanism $T(\xd) = C$ described above, where $C$ was designed to be the actual confidential dataset, the corresponding meta-mechanism is the identity function $T_{\mathrm{meta}}(\xd) = \xd$, whose PLB, infinity, reflects the true level of protection given by $T$. The meta-perspective therefore resolves the paradox of this 
perverse 
toy example. But real world examples often have complex data-dependent pipelines that are difficult to fully `algorithmize' into meta-mechanisms, %
making a true implementation of ``end-to-end'' DP challenging \citep[see e.g.][]{drechslerComplexitiesDifferentialPrivacy2024}.

\section{Highlights of Part II: The US Census's Evolving Data Protection}\label{secCensus}%

The Decennial Census of Population and Housing is a critical piece of US infrastructure. It determines the apportionment of seats in the House of Representatives; it is relied upon for allocating trillions of dollars in federal funding each year; and it informs the decision making of businesses, urban planners and hospitals, amongst many others \citep{bureauUsesDecennialCensus, reamerBrief7Comprehensive2020, ModernizingUSCensus1995}. Safeguarding such an important data source requires robust SDC, a task that the USCB has long taken seriously. In fact, the bureau has been managing the risk of indirect disclosure in the Decennial Census from 1940 onward \citep{HistoryCensusPrivacy}. In the 1990 Census, protections were strengthened and a new SDC method was introduced: \textit{data swapping} \citep{mckennaDisclosureAvoidanceTechniques2018, daleniusDataswappingTechniqueDisclosure1982,fienbergDataSwappingVariations2004}.

Data swapping (or record swapping) is a general concept that encompasses a broad class of algorithms. These algorithms select a set of records and then shuffle the values of certain variables among these records. We call the variables whose values are shuffled the \textit{swapping variables}; all other variables are called the \textit{holding variables}. Usually, records are partitioned into groups (or strata) according to the values they take on a subset of the holding variables we call the \textit{matching variables}; and records are only shuffled within their matching group.

The primary SDC methods used in the 1990, 2000 and 2010 Censuses were forms of data swapping, the full technical details of which have not been made public due to confidentiality concerns. We know however that the USCB swapped entire households, rather than shuffling person-level data between households; that the swapping variable was geographic (e.g., block group, tract, or county); and that the matching variables included broader levels of geographies (i.e., tract, county or state) as well as the household's total counts of adults and children. Furthermore, unique or unusual households that the bureau believed had higher disclosure risk had a higher chance of being swapped. (For references to these statements, as well as others in this section, see Part~II.)

In the 2010s, spurred by an increasing awareness of privacy risks in statistical products \citep{dinur2003revealing}, the USCB conducted a \textit{reconstruction attack} on the 2010 Census \citep{abowd2010CensusConfidentiality2023}. Using published tables and publicly available information on the relationships between these tables, they were able to determine with a high degree of confidence much of the underlying post-swapped microdata that produced these tables. This lead to a revolution at the bureau---the 2020 Census would not be protected by using data swapping, as was the case for the previous three decades, but rather by brand new SDC methods that were explicitly designed to satisfy DP \citep{abowd2018us}. Yet, does the USCB's official adoption of DP, on its own, truly represent a sea change in how it protects the census? %
What if, in fact, the census was already protected in 2010 by a DP method---or at least by a method that is very similar to a DP one?

\subsection{A DP Guarantee for the 2010 Census?}\label{sec:dp2010}

While data swapping was not originally invented with DP in mind, it may still be possible for it to satisfy some DP specification. 
However, it is effectively a method for adding noise \textit{only} to the relationship between the swapping and holding variables within each matching group. %
As such, the marginal distributions of these three sets of variables are invariant under swapping, as is the joint distribution of the swapping and matching variables. Data swapping can therefore only be DP subject to the invariants it induces---i.e., it can satisfy a DP specification only if the specification's data multiverse respects its invariants. This limitation on the scope of protection greatly reduces the SDC guarantee provided by DP. %

Moreover, some of the technical implementation details of the 2010 swapping procedure preclude it from satisfying a pure $\epsilon$-DP specification (i.e., a specification whose output premetric $\dPr$ is the same as the one used in the original DP specification of \citeauthor{dwork2006calibrating}, \citeyear{dwork2006calibrating}). 
Nevertheless, the \textit{Permutation Swapping Algorithm} (PSA)---which keeps to the spirit of the 2010 procedure, if not the exact implementation%
---does satisfy pure $\epsilon$-DP subject to the invariants it induces. 

The PSA is very simple to describe: It selects records independently with probability $p$ (the `swap rate') and then permutes the values of those records' swapping variables.\footnote{Incidentally, this idea, under the name $n$-Cycle swapping, was under active investigation by the USCB up until the bureau redirected its research efforts towards DP \citep{mckennaLegacyTechniquesCurrent2019}.}
The following statement is an informal version of the main result of Part~II, which proves that the PSA satisfies a DP specification.

\noindent
\textbf{Theorem~II.1} (informally)\textbf{.}  \textit{Subject to the invariants induced by it, the PSA %
satisfies pure $\epsilon$-DP, with
		\begin{equation*}%
            \epsilon \le \begin{cases}
			\ln (b + 1) - \ln o & \mathrm{if\ } 0 < p \le 0.5, \\
			\max \big\{ \ln o, \ln (b+1) - \ln o \big\} & \mathrm{if\ } 0.5 < p < 1,
		\end{cases}
        \end{equation*}
where $o=p/(1-p)$, and $b$ is the size of the largest matching group.}

In nontechnical terms, the first four building blocks of the PSA's DP specification are as follows. Firstly, the protection domain $\mathcal X$ is any set of datasets that all share the same variables. Secondly, what ``subject to the invariants'' means is that %
among all possible datasets in $\mathcal X$, %
the only ones that carry the protection guarantee are the ones
that share the same invariant values with the actualized, confidential dataset. This is because other datasets will be excluded from consideration by any competent attacker due to the fact that their invariant values are different to those published. %
Mathematically, the invariants partition the domain $\mathcal X$ into universes, with all datasets in each universe sharing the same invariant values. By ``subjecting DP to the PSA's invariants,'' we mean that DP's Lipschitz condition (see Equation~\ref{eqSec2Derivertive}) is restricted to datasets in the same universe.
Hence, comparisons between datasets with different values of the invariants are excluded from consideration.

Thirdly, the granularity of the PSA's DP specification is equal to the resolution of the PSA's swaps. For example, %
if the PSA swaps person records, then the granularity of protection is person records. More technically, we mean that the PSA's input premetric $\dX$ is the Hamming distance on person records. Thus, if a number $n$ of person records are changed, the distribution of the PSA's output will change by at most $\epsilon n$ units, as measured by the output premetric of the PSA's DP specification (assuming that the changes to these records do not result in a change in the value of any of the invariants). Fourthly, this output premetric is given by the maximum likelihood ratio---that is, the maximum value, over all possible outputs $t$, of the relative likelihood of observing output $t$, under input dataset $\xd$ as compared to under the some alternative input $\xd'$. 
Hence, the output premetric of the PSA's specification %
corresponds to the notion of pure $\epsilon$-DP.

Part~II then instantiates the PSA to align as closely as possible with what we know about the 2010 swapping procedure. %
This gives us a concrete DP specification reflecting the SDC protection afforded to the 2010 Census---although with the caveat that this assumes
this census was protected by the PSA. As such, we emphasize that we do not provide a DP specification for the actual 2010 Census since such a specification must reckon with the exact implementation details of the 2010 SDC methods, not the PSA. However, because we believe the PSA can closely parallel the 2010 swapping procedure by appropriately choosing its implementation parameters, the resulting DP specification is nevertheless a useful perspective on the protection provided to the 2010 Census.

To mimic the 2010 swapping procedure, the choices for the PSA's implementation parameters are as follows. The 2010 DAS was integrated into the census data pipeline after all imputation and editing processes were completed; 
we place the PSA at the same place in the pipeline. This means the PSA's protection domain is the set $\Xcef$ of all possible \textit{Census Edited Files}---i.e., all hypothetical outputs resulting from the first stages of the census data pipeline through to the imputation and editing processes. This has important implications on what data is actually being protected by the PSA: the edited and imputed records, not the `raw' census responses. Moreover, because $\Xcef$ determines what it means to counterfactually alter data, it also has implications on what the protection unit in 2010 was. Indeed, mirroring the 2010 swapping procedure, we set the PSA to swap household-level records, so that its input premetric $\dX$ is the Hamming distance on household records; yet this does not imply the 2010 protection units were households---because a change in a single record of the Census Edited File does not always correspond to a single household changing their census responses. Instead, the 2010 protection units are `post-imputation households'---imaginary entities that can alter their own records in the Census Edited File freely without affecting other, imputed records. (See Part~II for an explanation of how the PLB must be inflated in order to have households as the protection unit.) 

We set the PSA's matching variables to be the household's state and size, and its swapping variables to be the household's county. This results in a multiverse $\scD_{2010}$ where all statistics at the state and national levels are invariant, as well as the counts of households by size at the block level. Finally, we set the swap rate $p$ to be 2--4\%, as \citet{boyd2022differential} state was used in 2010. This results in a PLB $\epsilon$ between 18.29 and 19. %
However, since the 2010 swapping procedure also included the number of voting-aged people as a matching variable, 18.29--19 is an upper bound for 2010's approximate PLB. (%
We cannot compute the PLB when the number of adults is invariant because the necessary statistic---the value of $b$ in Theorem~II.1---is not publicly available.) However, this also implies that $\scD_{2010}$ gives a lower bound on the invariants: %
in addition to the statistics reported above, the block-level breakdown of households by the number of adult occupants (and hence also by the number of children occupants) was invariant.

\subsection{The DP Guarantee of the 2020 Census}\label{sec2020Guarantee}

Having established that the 2010 Census may be analyzed from the perspective of DP, it is fruitful to compare it with the DP specification of the 2020 Census. We  start this comparison by examining each of the five DP building blocks in turn. As in 2010, the 2020 disclosure avoidance system (DAS) also took the Census Edited File as input. Hence, the protection domain remains constant across the two census; in both cases it is the set of all possible Census Edited Files $\Xcef$. Secondly, the granularity of protection in 2020 was person-level records. All other components being equal, this would imply weaker protection than in 2010, in which household-level records were directly protected; but, as we will see, the three remaining components are not equal. 

Most importantly, the 2020 DAS has far fewer invariants than was the case in 2010. The 2020 invariants were carefully considered and minimized to those required by operational and constitutional mandates. These invariants are the state populations as well as the counts at the block level of housing units and of each type of occupied group quarters. %
Initial analysis suggests that these invariants have minimal effect on SDC; the same cannot be said about 2010's invariants \citep{abowd2010CensusConfidentiality2023}.

The output premetric (i.e., `standard of protection') used in 2020 is what we call the \textit{normalized R\'enyi metric}; this corresponds to the notion of zero-concentrated DP (zCDP) \citep{bunConcentratedDifferentialPrivacy2016a}. Under these settings for the first four components, the 2020 Census's PLB is given by $\rho^2 = 55.371$. (We follow the standard convention of using $\rho$ to denote the PLB in the context of zCDP.) %
Additionally, we may translate from zCDP to $(\epsilon, \delta)$-DP and thereby also express the 2020 PLB by $\epsilon = 126.78$ with $\delta = 10^{-10}$. (To be clear, in this translation across DP specifications, the first three building blocks stay the same, while the output premetric changes from the normalized R\'enyi metric [defined in Part~I] to the $\delta$-approximate multiplicative divergence, defined in Equation~\ref{eq:delta}.) %

It is worth noting that the above DP specification only assesses the disclosure risk associated with the primary 2020 Census products. As of the time of writing, there have already been other releases of the 2020 Census data, and more are planned for the future.%
\footnote{The primary 2020 Census data products are the P.L. 94-171 Redistricting Summary File \citep{uscensusbureauPLnational2021,uscensusbureauPLstate2021}, the Demographic and Housing Characteristics (DHC) File \citep{uscensusbureauDHC2023}, the Detailed DHC-A and -B Files \citep{uscensusbureauDDHCA2023, uscensusbureauDDHCB2024}, the Supplemental DHC \citep{uscensusbureauSDHC2024} and related auxiliary products. Examples of future releases that rely on the 2020 Census include the annual Population and Housing Unit Estimates and the National Population Projections \citep{uscensusbureauMethodologyUnitedStates2023, uscensusbureauMethodologyAssumptionsInputs2023}. See Part~II for more details.}  %
While we have not been able to obtain information on the SDC of these releases, they will necessarily increase the 2020 PLB and may possibly weaken the other components of the 2020 DP specification. A degradation of the DP specification due to additional data releases is not a concern under data swapping; the DP specification from the previous section covers all the 2010 Census products because they were all generated from the post-swapped microdata (see the end of Section~\ref{secAdvantages}).

Beyond the core details of the 2020 DP specification explained above, several other aspects merit attention. Firstly, as in 2010, setting the protection domain to be $\Xcef$ in 2020 has important implications: The 2020 Census does not have ``end-to-end'' DP protection \citep[c.f.,][]{huProvablePrivacyNonprivate2024a}; its protection units are `post-imputation persons' and as such it does not provide protection to individuals' census responses directly. Secondly, a more nuanced perspective on the 2020 DAS would examine its \textit{per-attribute} PLBs \citep{ashmead2019effective}. A per-attribute analysis considers a DP specification in which only one variable (i.e., attribute) is allowed to vary within each data universe. This allows for a finer-grained assessment of SDC, rather than assuming the worst-case possibility of complete dependence between variables when composing the per-attribute budgets into a single total budget. Apart from the following two observations, we leave this important discussion to future work: The per-attribute budgets are much smaller than the overall 2020 PLB; and a per-attribute analysis is not applicable to data swapping since its DP specification does not rely on the composition theorem of DP.

\section{New Conundrums in Familiar Trade-Offs}\label{secConundrums}
\subsection{Traditional SDC and DP: Can We Reap the Benefits of Both?}\label{secAdvantages}

We have demonstrated that data swapping, being a quintessential example of traditional SDC, can be analyzed under our system of DP specifications. This analysis enables a comparison of the SDC protections provided in the 2010 and 2020 Censuses, sketching an evolution of the USCB's approach to disclosure avoidance. Meanwhile, it serves as a proof-of-concept that a data protection method, like swapping, can enjoy some of the distinctive advantages of DP without forgoing the strengths of traditional SDC.
Other work have examined other traditional SDC methods through the lens of DP to varying degrees of success
\citep{rinottConfidentialityDifferentialPrivacy2018, bailieABSPerturbationMethodology2019, sadeghiConnectionABSPerturbation2023, neunhoefferFormalPrivacyGuarantees}.
These efforts stem from a shared acknowledgment that both approaches have their own unique advantages \citep{slavkovicStatisticalDataPrivacy2023}, and continuing to bridge the gap between these two paradigms %
may bestow opportunities to reap the best of both worlds.

We are reminded, however, that DP and traditional SDC are grounded in distinct design philosophies, which drive the two approaches to place emphasis on different aspects of the data protection method. Some of these distinctions are highlighted by the DP specification framework that we put forth.  On one hand, a DP specification gives a precise, mathematical formulation of the SDC it provides. Take the PSA for example: its DP specification delimits the information the PSA does not protect (its invariants) and the extent to which it protects the remaining information. %
It describes the granularity at which attackers are limited in learning about aspects of the confidential data that are not disclosed by the invariants alone, and the standard against which this learning is measured. In short, the PSA's DP specification answers the `who', `where', `what', `how' and `how much' questions of protection. These answers serve as a useful aid in assessing whether the PSA is appropriate for a given data release, and, if so, in choosing its implementation parameters.

While this chapter has largely focused on its ability to describe SDC protection, DP is not just a descriptive framework. It also provides a calculus for reasoning about how protection loss accumulates across multiple data products, a tool which is becoming increasingly more valuable as national statistical offices diversify their offerings \citep{kitchinOpportunitiesChallengesRisks2015}. Moreover, adopting a DP flavor as the yardstick for measuring SDC permits complete transparency of the data release mechanism, since DP guarantees do not degrade with the attacker's \textit{knowledge of the mechanism} (in contrast, degradation can occur with the attacker's knowledge of the relationships among the data subjects). %
For example, even when armed with the complete knowledge of the PSA's implementation details (including the values of all its parameters, such as the swap rate or swap key, but excluding, of course, the value of its random seed), it is still impossible for an attacker to thwart the SDC protections described by its DP specification. While transparency does not assist attackers in breaking DP's protection guarantees, it is important to legitimate data users as an essential prerequisite for valid statistical analysis of privacy-protected data \citep{gongTransparentPrivacyPrincipled2022}. Indeed, in Section~\ref{secTransparency} we explain how, by allowing quantitative analysts such as statisticians 
and social scientists to correct %
for the statistical errors induced by SDC protection, transparency increases data utility and supports robust research findings. %

On the other hand, traditional SDC techniques enjoy design advantages that most DP methods do not. For example, data swapping maintains \textit{facial validity}---the 2010 Census outputs all look `reasonable' in the sense that there are no negative or fractional counts, nor are there implausibly large or small reported values. %
More generally, the 2010 publications pass the sanity checks an observant reader might make, which is useful for building trust in the census among the general public. From the opposite perspective, a lack of facial validity is an important concern for statistical agencies like the USCB. It presents an issue for data users who are confused and disinclined to use seemingly erroneous data; it erodes the public image of the agency; and it hampers efforts to improve differential response rates among disadvantaged communities \citep{boyd2022differential, drechslerDifferentialPrivacyGovernment2023, Oberski2020Differential}.\footnote{Related, but distinct, to facial validity is the concept of \textit{face privacy}, which is the requirement that the data release mechanism produce output that appears to the casual observer to offer privacy protection \citep{hodDifferentiallyPrivateRelease2024}. In contrast, facial validity requires that the output is a plausible representation of the real world, even to a data user who is unaware that artificial noise was added for SDC protection.}

\textit{Logical consistency} is another advantageous property of data swapping, as it ensures all statistics produced from the 2010 Census align with one another. For example, in any contingency table released in 2010, the sum of cells in a single row or column always matches the marginal total. Likewise, reported values for the same count remain consistent across different publications. Additionally, the 2010 Census outputs respect the structural zeroes and edit constraints present in the underlying confidential microdata. %
In contrast, many of the 2020 Census outputs will not be consistent across publications and even within the same publication, row- and column-sums will not match the reported totals. %
(However, the outputs produced by the TDA---the PL and DHS files---are logically consistent, although as we will see, this comes at a cost.)

Like facial validity, logical consistency is important for users and advocates of census data \citep{boyd2022differential, hotz2022chronicle, rugglesDifferentialPrivacyCensus2019}. 
Yet, many DP methods do not maintain facial validity nor logical consistency, and others, such as the TDA, cannot satisfy these properties without partially destroying statistical transparency.\footnote{A data release mechanism $T$ is statistically (or probabilistically) transparent if the conditional probability distributions $\sfPx (T \in \cdot)$ are public knowledge \citep{gongTransparentPrivacyPrincipled2022}. Statistical transparency is distinct to algorithmic transparency, which requires that the source code of the mechanism $T$ is disseminated. For all practical purposes, statistical transparency is a stronger requirement than algorithmic transparency since $T$'s source code may be so complex that it is practically impossible to derive the conditional distribution it induces.} This is because the majority of DP methods rely on optimization-based post-processing to restore facial validity and logical consistency \citep[e.g.][]{barak2007privacy,hay2010boosting}. 
Optimization-based post-processing can be algorithmically transparent but in most cases it destroys the statistical transparency of the resulting two-step privacy mechanism---a crucial requirement for principled statistical analysis \citep{gongTransparentPrivacyPrincipled2022}. %
The recent proposal by \citet{dharangutte2022integer} does away the need for post-processing when the noise infusion is additive. %
However, it relies on MCMC sampling and hence is nontrivial to implement for large-scale data products. In contrast, swapping achieves facial validity and logical consistency automatically without the need for additional computation. %

Furthermore, data swapping---like other traditional SDC techniques but unlike many DP methods (such as those used in 2020)---is easy to communicate and understand at a high level by a broad, nontechnical audience. This is important for building trust and maintaining the buy-in of data providers, custodians and other stakeholders. 
Swapping is also easily implementable and amenable to the types of data collected by government agencies, as evidenced by its use in the US, the UK and the EU \citep{mckennaDisclosureAvoidanceTechniques2018, officefornationalstatisticsonsProtectingPersonalData2023, devriesOverviewUsedMethods2023}. 

Finally, as a pre-tabular perturbation method, swapping also has the advantage of producing a `synthetic' dataset that serves as the source for all census publications. %
This simplifies the data release process as all outputs are derived from this `post-swapped' data without requiring additional SDC treatment. %
This also explains why the 2010 publications maintain logical consistency; because no further noise is introduced, all outputs are consistent with the post-swapped data and hence also with each other. %
Moreover, this approach ensures that the release of additional data products in the future not only maintains \textit{backwards compatibility} with previous products, but also does not degrade their DP specification. 
Indeed, as long as all products are based on the same post-swapped microdata, they are all covered under the DP specification of the swapping algorithm that produced this microdata (by DP's post-processing theorem). %
In this way, data swapping allows the statistical agency to publish a single DP specification that encompasses all existing and future publications. %
As mentioned in Section~\ref{sec2020Guarantee}, this stands in contrast to the 2020 Census. Each publication from the 2020 Census has its own DP specification, and to understand the SDC provided to the census data as a whole, one must aggregate these DP specifications into a single comprehensive one---a process that must be repeated with each new publication. And, because every data product introduces additional disclosure risk, the overall 2020 DP specification weakens after each new release, in comparison to the single, upfront DP specification associated with data swapping.

\subsection{Between Transparency and Epistemic Uncertainty}\label{secTransparency}

Another fundamental difference between traditional SDC and DP is their attitude toward transparency, that is, the publication of the details of the data protection method.
A common principle in traditional SDC is that privacy requires some degree of obscurity. A data custodian should not fully reveal the implementation details of their SDC method, because these details could be used by an attacker to unpick the method's SDC protection (see \cite{slavkovicStatisticalDataPrivacy2023} and references therein). 
As such, the privacy protection offered by any randomized SDC algorithm is not solely due to the aleatoric uncertainty associated with the noise it injects, but also due to the epistemic uncertainty arising from the attacker's deficient knowledge of the algorithm. %
The inherent randomness of the SDC method provides direct protection, and 
the plausible deniability surrounding how exactly the method was implemented provides indirect protection. However, epistemic uncertainty is unfortunately harder than its aleatoric counterpart to model and reason about; see for example the extensive literature on imprecise probabilities (IP) \citep{shafer1976mathematical, walleyStatisticalReasoningImprecise1991, augustinIntroductionImpreciseProbabilities2014}. Consequently, the privacy protection provided by the epistemic uncertainty of an SDC method is difficult to describe with mathematical guarantees and, as far as we are aware, has not been systemically studied. Nevertheless, the principle of `privacy through obscurity' is frequently invoked by national statistical offices in reference to their SDC methods \citep{mckennaDisclosureAvoidanceTechniques2018, ukstatisticsauthorityTransparencySDCMethods2021, chipperfieldAustralianBureauStatistics2016}. 

Establishing a DP guarantee for a traditional SDC method provides a rationale for reducing obscurity: as the DP guarantee does not degrade with knowledge of the mechanism (as explained in Section~\ref{secAdvantages}), %
the data custodian is justified in %
being publicly transparent 
about the SDC method. This would provide crucial insights into the quality of existing data products in ways that were not previously possible.
For example, data swapping is not traditionally a transparent SDC technique; in fact, the currently available public documentation on the 2010 DAS is deliberately deficient, stymieing researchers' ability to appropriately account for the noise it injects into census data \citep{kennyEvaluatingBiasNoise2023}. 
The explicit statement of the DP specification of the PSA provides a theoretical justification for releasing the implementation details of the 2010 swapping procedure. Indeed, the publication of a complete description of the 2010 DAS, as the USCB has done for the 2020 DAS, would greatly facilitate any comparative analysis at the heart of the ongoing debates about the trade-off between SDC and data utility in the census.

There are more reasons to %
publish the algorithmic specification of SDC procedures like swapping. Despite being the main SDC method for the census in 1990--2010, it remains difficult to fully quantify swapping's adverse impact on data quality. A peek into the technical specification of swapping can add tremendous utility to the data products it protects. 
Data users who conduct complex statistical modeling with official data products may particularly benefit from the transparent knowledge of the SDC methods used in creating those products. For example, it is well understood in the literature that data swapping negatively affects the quality of downstream data analyses. Specifically, because swapping adds noise to the relationships between the swapping and holding variables, it reduces researchers' ability to study these relationships. %
\citet{mitra2006adjusting} and \citet{drechslerSamplingSynthesisNew2010} demonstrate that even low swap rates (e.g., 5\%) can substantially reduce the effective  coverage of confidence intervals for the regression coefficient between swapping and holding variables. Such deterioration in coverage is in no small part due to performing na\"ive statistical analyses that do not account for the SDC mechanism. %
As \citet{gongTransparentPrivacyPrincipled2022} demonstrates, performing standard regression analyses on data protected via DP noise infusion results in similar types of coverage deterioration. Furthermore, this deterioration can be restored once the DP method is statistically modeled---at the expense of wider, though valid, intervals. With that said, an analyst cannot be blamed for performing na\"ive analyses on swapped data when details of the swapping procedure are not public knowledge.

Publishing the details of the USCB's swapping methods will also help to dispel the ``statistical imaginaries''  associated with the censuses of 1990--2010 \citep{boyd2022differential, sarathyStatisticalImaginariesState2024}. It is easier to argue that census data contain errors due to SDC protection when one can point precisely to how these errors were introduced and quantify their distribution exactly. Transparency enables concrete statements like, `the expected error due to SDC protection is $x$\%,' in place of vague expressions such as, `these counts may have some error as their contributing households could have been swapped.' In general, raising awareness about an official data product's SDC errors by publicly documenting the methods used to protect them is a step towards ``shifting the statistical imaginary to account for uncertainty'' \citep{boyd2022differential}, thereby improving the legitimacy of the data product and the perception of accountability of the data custodian. Such a paradigm shift occurred in 2020 due to the USCB's transparency about their new SDC methods; a similar shift in the imaginary of earlier census data could be sparked by a comparable level of transparency about  
their 2010 SDC method. %

However, we caution that the idea a DP method can be transparently disseminated at no cost to privacy rests on two premises. The first is that the epistemic uncertainty of an SDC method is not a legitimate form of privacy protection. DP follows an intellectual tradition originating in cryptography that dismisses epistemic uncertainty as a brittle form of protection---brittle because it depends on a watertight security system to safeguard an SDC method's implementation details. Indeed, any measure of SDC protection that relies on epistemic uncertainty must depend on assumptions about the attacker's knowledge of the data release mechanism; therefore, epistemic-based SDC can degrade as an attacker learns new information. As such, 
existing measures of protection found in the DP literature---i.e., DP specifications---are 
solely based on the aleatoric uncertainty arising from an SDC method's stochasticity, %
justifying DP's tenet of `transparency for free.' Nevertheless, our system of DP specifications could in principle be extended to also incorporate protection based on epistemic uncertainty.

Towards this end, one open research direction 
is to leverage machinery from IP to quantify the epistemic uncertainty due to partial knowledge of the data release mechanism.
If, instead of full statistical transparency of the data release mechanism (i.e., public dissemination of the likelihood $\sfPx$), the data custodian could opt to publish an IP version of $\sfPx$. This would provide a level of transparency that is sufficient for principled statistical analysis because an analyst can correct for the noise introduced by the mechanism using IP tools. 
At the same time, an attacker's ability to infer confidential information would be reduced because they no longer have access to the precise $\sfPx$, but only to its IP version. This increase in protection could be incorporated into a DP specification through the use of an output premetric that quantifies the difference between two IP objects, rather than between two precise probabilities. Research along these lines will explore new Pareto frontiers in the trade-off between SDC and utility by introducing---and rigorously quantifying---a controlled amount of `privacy through obscurity' without losing the transparency required for statistical analysis.

The second premise is that the dissemination of a DP method's specification does not itself incur a cognizable privacy cost. This premise becomes questionable when the specification includes invariants. 
In the case of the PSA (or any other invariant-preserving DP mechanism), %
publishing its specification necessitates revealing the complete list of all its invariants. %
However, as we have repeatedly emphasized, there may be a disclosure risk associated with the knowledge of what swapping's invariants are, in and of itself---especially when its invariants are at a fine level of granularity. For example, the 2010 invariants are privacy revealing in an intuitive sense, even if they are not technically privacy revealing in the sense that their publication does not increase the PLB under the PSA's DP flavor. %
Indeed, too many invariants supply the attacker with confidence in their efforts to reconstruct the microdata and reidentify individual records. With many invariants present, the DP specification itself could carry a nontrivial amount of information, and some could consider that a breach of privacy, even though the sense of privacy here extends beyond the scope of the DP specification in question. 
We are sympathetic to this view and therefore advocate for careful deliberation, weighing the cost of making public the invariants (through releasing the DP specification) against the benefits of algorithmic transparency this allows.

\section{Conclusion: Can Swapping Be DP?}
\label{secConclusion}

In this chapter, we have taken a ``stirred, not shaken'' approach to two central subjects under study: DP and data swapping. Our goal is neither to revamp these notions nor to rob them of their essence. On the contrary, our overarching goal is to \textit{reveal} their essence, which is achieved 
by laying down a rigorous explication of DP through our five-building-block system, %
and by reexamining data swapping through the lens of this system. %
We want to be clear that we are not advocating to reverse the progress that the USCB and the data privacy research community have made to advance SDC. In fact, we seek to continue this progress by building stronger connections between DP and traditional SDC so as to reap the benefits of both fields. We believe that this approach ultimately facilitates the modernization of SDC in a way that is helpful to data custodians, responsible to data contributors, and respectful to data users.

Returning to the question raised in the introduction of this work---is swapping DP?---we observe that the term DP has been used to refer to multiple distinct notions, potentially leading to some understandable confusion on the part of the reader. 
Indeed, while summarizing an entire field of work is difficult, we identify four broad concept categories associated with DP. Firstly, we have primarily used the term DP to refer to the class of technical standards encapsulated by the Lipschitz condition in Equation~\ref{eqSec2Derivertive}, which we have variously called DP definitions, formulations, or specifications. Secondly, DP as a field also consists of %
its data release mechanisms, which can come in three different representations: %
as mathematical abstractions---useful for proving that these mechanisms satisfy their DP specifications; %
as algorithmic implementations---useful for employing these mechanisms to release data; %
and as sociotechnical deployments---useful for embedding these mechanisms into real world contexts. %
(See \cite{seemanPrivacyUtilityDifferential2023}, for a discussion on the distinctions between these representations of data release mechanisms, and why these distinctions matter.) Thirdly, DP can be considered as a framework by augmenting the first two concepts---DP specifications and mechanisms---with a set of tools for reasoning about these concepts (e.g., composition theorems or programming frameworks---see \cite{steinkeCompositionDifferentialPrivacy2022} and \cite{gaboardiProgrammingFrameworksDifferential2024}).

Finally, DP can be considered as a philosophy that formulates a particular way of thinking about data privacy \citep{mckaybowenPhilosophyDifferentialPrivacy2021, bunStatisticalInferenceNot2021, dwork2006differential-dalenius, seemanPrivacyUtilityDifferential2023}.
For our purposes, we focus on one aspect of this philosophy: By default, every attribute of the confidential dataset should be treated as sensitive, since we do not know how it may be used by an attacker in the future. As such, every statistic---even if noisy and only partially informative of the dataset's attributes---should be considered as privacy revealing.\footnote{However, there are exceptions to this rule. For example, the dataset size in bounded DP is not considered privacy revealing.}
In contrast, the philosophy behind data swapping, as explained in Part~II, is based on thwarting reidentification attacks by adding noise to the relationship between sensitive attributes and quasi-identifiers. Thus, swapping's philosophy does not conform with DP's because it only adds noise to one attribute of the data, not all of them. %

The misalignment between their philosophies---which is fundamentally a tension between invariant statistics and DP---is a major reason to conclude that swapping cannot be DP. 
However, as we argue in Section~\ref{secIntro}, a strict resistance to invariants %
greatly limits DP's applicability, since it does not allow for the possibility that some information cannot be protected---either because it is already public, or because it is a mandated invariant. It is our view that a good theory of SDC should not %
delineate what does and does not need protection. Rather, %
it should provide a flexible framework that enables practitioners and policymakers---who possess the necessary contextual information---to decide what level of protection should be afforded to each data attribute. Indeed, 
this perspective is supported by the broader literature \citep{kifer2014pufferfish, he2014blowfish}, even if it is misaligned with the default approach to DP.

Nonetheless, there is considerable alignment between the ideas behind DP and swapping. For one, they both agree on the tenet that SDC protection arises from epistemic uncertainty (see Section~\ref{secTransparency})---i.e., that sensitive attributes should be protected using randomization. Moreover, it was possible to mold a particular version of swapping---the PSA---to conform to the particular way epistemic uncertainty is measured by pure DP. This alignment was ultimately what allowed us to prove that the PSA satisfies a DP specification, placing it squarely within the framework of DP and answering the question of whether swapping can be DP in the affirmative.

\bookmarksetup{startatroot}
\phantomsection\subsection*{Acknowledgments} 
We thank Daniel Susser for his helpful and stimulating feedback on early drafts of this work, and Priyanka Nanayakkara for her proofreading of later drafts and valuable suggestions, particularly with regard to Figure~\ref{figDPDiagram}. We are grateful to the participants of the National Bureau of Economic Research's conference \textit{Data Privacy Protection and the Conduct of Applied Research: Methods, Approaches and Their Consequences} (May 4--5, 2023); the \nth{36} New England Statistical Symposium's invited session \textit{A Private Refreshment on Statistical Principles and Senses} (June 6, 2023); the 2023 Joint Statistical Meetings session \textit{Methodological Approaches to Privacy Concerns Across Multiple Domains} (August 10, 2023); the CA Census retreat at the Boston University Center for Computing and Data Science (September 26, 2023); and the Statistics Canada Methodology Seminar (October 31, 2023) for their thoughtful comments and questions. We appreciate enormously the detailed feedback provided by Daniel Kifer, John Abowd, Philip Leclerc, Ryan Cummings, Rolando Rodriguez, Robert Ashmead, Sallie Keller and Michael Hawes at the US Census Bureau, which greatly improved and corrected all three parts of this trio of articles. All remaining errors are purely our own. JB gratefully acknowledges partial financial support from the Australian-American Fulbright Commission and the Kinghorn Foundation; RG and XLM acknowledge partial financial support from the NSF; and XLM acknowledges partial financial support from Harvard University's Office of the Vice Provost for Research.

\bookmarksetup{startatroot}
\begingroup
\sloppy
\phantomsection\printbibliography
\endgroup

\end{document}